\documentclass[%
 reprint,
 amsmath,amssymb,
aps,
]{revtex4-1}
\usepackage{color}
\usepackage{subfigure}
\usepackage{graphicx}% Include figure files
\usepackage{dcolumn}% Align table columns on decimal point
\usepackage{bm}% bold math
\usepackage{hyperref}% add hypertext capabilities
\usepackage{soul}
\usepackage{amsmath}
\usepackage{lipsum}
\begin{document}

\preprint{APS/123-QED}

\title{Time-modulated Measurements: Material Characterization Based on Scalar Reflection Data}% Force line breaks with \\

\author{Sepideh Ghasemi$^{1}$, Hamid Rajabalipanah$^{2}$, Majid Tayarani$^1$, Ali Abdolali$^2$, and Mohammad Baharian$^3$}
\affiliation{$^1$School of Electrical Engineering, Iran University of Science and Technology (IUST), Tehran, Iran}
\affiliation{$^2$Applied Electromagnetic Laboratory, School of Electrical Engineering, Iran University of Science and Technology (IUST), Tehran, Iran}
\affiliation{$^3$School of Electrical and Computer Engineering, University of Tehran, Tehran, Iran}


%\affiliation{
%\affiliation{
 %Third institution, the second for Charlie Author
%\affiliation{%



\begin{abstract}
This paper explores the time-modulated waveguide setups for unique and accurate permittivity and permeability extraction of lossy dispersive samples with phase-less measurements. We theoretically demonstrate that when the position of the short-circuit termination is dynamically modulated in a predefined way, the phase information of the reflection S-parameter manifests itself into the amplitude level of the emerging harmonics. Being insensitive to the calibration plane shifts and phase uncertainties in reflection measurements while bypassing a priori knowledge about the material under test (MUT) and also the transmission coefficient, can be enumerated as the main advantages of the proposed time-modulated retrieval scheme. Moreover, the presented reconstruction algorithm offers a simple post-processing step to facilitate fast computations of $\varepsilon_r$ and $\mu_r$. Several illustrative examples at X-band frequencies have been presented to numerically verify the validity of the proposed approach for some applicable and practical types of homogeneous materials. Two possible realization and measurement configurations are suggested and discussed based on mechanical actuation and electrical phase control. We have also performed an uncertainty analysis to examine how the realization tolerances can affect the accuracy of results. By involving the temporal dimension, the proposed strategy takes a great step forwards in phase-less reconstruction of the electromagnetic parameters.
%\begin{description}
%use the optional argument of the \verb+\item+ command to give the category of %each item. 
%\end{description}
\end{abstract}

\pacs{Valid PACS appear here}% PACS, the Physics and Astronomy
\maketitle

%\tableofcontents

\section{INTRODUCTION}
Electromagnetic characterization of materials is a reach, huge and old topic of microwave engineering. Almost any application of electromagnetic energy in which electromagnetic wave interacts with some sort of material, requires precise knowledge of complex permittivity and permeability. Electromagnetic characterization is used in a diverse field of scientific and industrial applications such as agriculture \cite{Nelson}, bioengineering \cite{Shimonov}, communication \cite{Seo}, imaging \cite{bahar1}, optical processing \cite{babilov, momeni}, and military systems \cite{Dester}. 

Microwave techniques proposed for material characterization can be categorized as resonant and non-resonant methods. Resonant methods are highly accurate but suffer from being narrowband, and material properties are just inferred in some discrete frequency points.  Non-resonant methods, on the other hand, are broadband with acceptable accuracy and can be used for a wide class of materials such as composites, very lossy samples, and metamaterials \cite{Hasar2010}. 

In one classification, non-resonant methods can be grouped into transmission/reflection and reflection-only techniques. Transmission/reflection is the conventional material characterization technique in which material under test (MUT) is loaded in a specific transmission line (such as coaxial line or rectangular waveguide). Complex permittivity and permeability are determined by measuring and processing the transmission and reflection data \cite{Nicolson, Weir, Baker-Jarvis, Boughriet}. Although transmission/reflection methods have been widely investigated for material characterization, they have some restrictions, and specific issues should be handled. For example, some artifacts arise in the extracted material parameters at frequencies near half-wavelength resonances for low loss material \cite{Qi}. On the other hand, very lossy materials may trap and dissipate the impinging wave, which results in weak and low SNR transmitted signal \cite{Hasar2014}, degrading the extracted material parameters. Beside the transmission/reflection techniques, the reflection-only methods have been also reported to extract the MUT parameters by performing only reflection measurements. These techniques are of special interest due to the facilitation of measurement setup and the ability to retrieve the material parameters when transmission data is not available or is very noisy. This is the case for conductor-backed and radar absorbing materials \cite{Fenner}, \cite{Hyde}. Reflection-only configurations require simpler microwave components, which reduces the overall cost of the measurement setup.

Another important feature of the non-resonant measurement approaches is the nature of the measured quantity. Generally, conventional methods read complex scattering parameters to characterize the MUT \cite{Weir}. By the way, it has been suggested to exploit amplitude-only data by accomplishing the scalar measurement of scattering parameter for the material characterization \cite{Hasar2010}. Although scalar measurements, when compared to complex measurements, introduces some theoretical challenges in the extraction process, they have some remarkable advantages. Since the conventional parameter retrieval strategies are significantly affected by the reference-plane at which the phase data is captured, the complex measurement techniques require accurate calibration kits for reading the phase accurately. This requirement is quite relaxed in the case of scalar measurements. Meanwhile, microwave systems accompanied with the amplitude measurements are much simpler, resulting in a more cost-effective system. 
Various reflection-only and/or scalar methods have been suggested in the literature so far \cite{Fenner,Hasar2017,Hyde2014,CatalaCivera, Huang2001}. Here, some more important studies are reviewed. Hassar \textit{et. al.} \cite{Hasar2017} introduced an elegant phase-less measurement technique for extracting permittivity of MUT. Nevertheless, in addition to supporting only non-magnetic materials, this method is restricted to low-materials as it resorts to both transmission and reflection measurements. A new method based on amplitude-only has been developed in \cite{Hasar2010} for low-loss materials backed by a short-circuit termination. However, magnetic and dispersive materials have not been studied. Hyde \textit{et. al.} \cite{Hyde2014} suggests a system for extracting complex permittivity and permeability of MUT from reflection-only measurements. Although interesting, this technique is not scalar. In \cite{Fenner}, a phase-less reflection-only technique is proposed to extract complex permittivity of MUT. Although this technique is very appealing, it is restricted to non-magnetic materials. \\ \\
As seen above, there is no phase-less reflection-only technique that provides complex permittivity and permeability with broadband specifications. The aim of this paper is to suggest a scalar reflection-only method for complex material characterization by using the benefits of time-varying systems.  Time-varying media and time modulated electromagnetic systems have drawn great attention of electromagnetic society in recent years \cite{Taravati,Pach,Caloz,Reyes,Liu2019,Sounas2017,Ptitcyn, Rajabalipanahk, Rajabalipanahf, Rajabalipanahc}. However, the concept of time-varying system in a characterization system has not been addressed, yet. To the authors' best knowledge, this is the first time that the concept of time-varying electromagnetic systems is explored to provide the required information for inversion in a phase-less problem. The idea is to place the MUT in a waveguide with a time-modulated short circuit termination. A probing electromagnetic wave with angular frequency $\omega_0$ illuminates MUT while at the same time, the short circuit is moving according to a specific pattern. Through adding the change over the temporal dimension, the reflected wave from MUT contains frequencies different than  $\omega_0$, in which the additional information is preserved in their amplitude. Our theoretical formulation proves that the amplitude information of the emerging harmonics can be elaborately employed to extract the phase information of the reflection S-parameter required for material characterization. We demonstrate several illustrative examples to verify the validity of the proposed phase-less approach in broadband reconstruction of $\varepsilon_r\left(\omega\right)$=$\varepsilon_r^\prime\left(\omega\right)-j\varepsilon_r^{\prime\prime}\left(\omega\right)$ and $\mu_r\left(\omega\right)$=$\mu_r^\prime\left(\omega\right)-j\mu_r^{\prime\prime}\left(\omega\right)$. Also, the possible realization schemes are comprehensively discussed to prove the feasibility of study. We believe the notion of time modulation, although at its early stages of maturity, can pave the way for future efficient, low cost and accurate characterization systems.
\begin{figure*}[t]
	\includegraphics[scale=0.35]{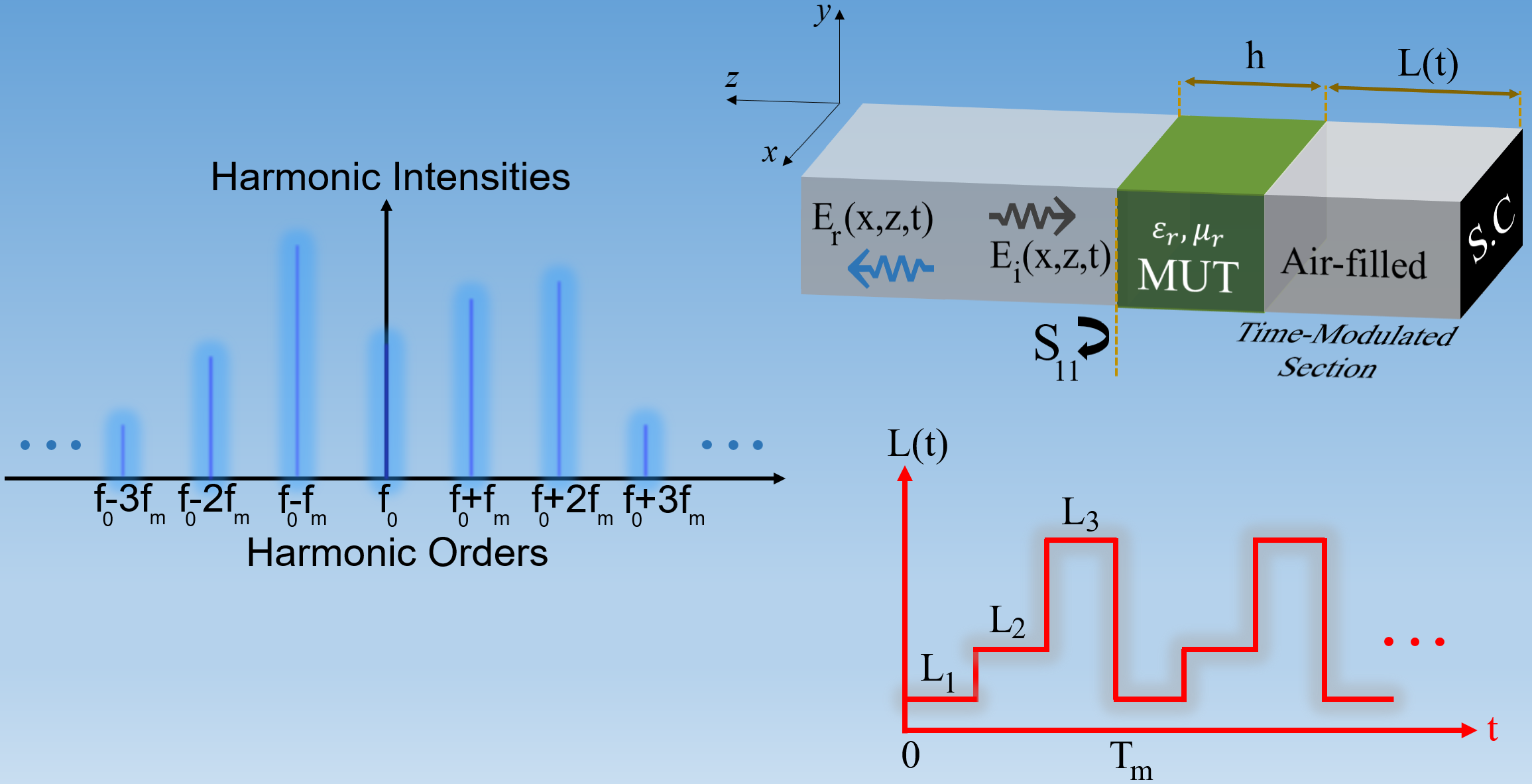}
	\caption{\label{fig:epsart} {Conceptual illustration of time-modulated waveguides for scalar material characterization of homogeneous samples. Temporal modulation of the short-circuit termination yields an infinite set of nonlinear harmonics carrying the information of reflection phases.  }}
\end{figure*}

\section{Theoretical Analysis }
\subsection{Background: Time-modulated Waveguides}
Prior to presenting the inverse problem, we investigate the interaction of the electromagnetic fields with a dielectric sample loaded in a time-varying waveguide. \textcolor{blue}{Fig. 1} illustrates the schematic of a time-modulated waveguide measurement system wherein the material under test (MUT) with known thickness  \textquotedblleft h\textquotedblright~is located at a distance from the short-circuit termination which is changing over time. Multiple-frequency S-parameter measurements \cite{Hasar2014li},
spectroscopic ellipsometry \cite{Fuji}, wavelength scanning \cite{Coskun}, spectral reflectometry/interferometry \cite{Lunacek}, and full spectra fitting techniques \cite{Huasong} can be utilized for determining the sample thickness. Without loss of generality, the MUT is considered to be homogeneous, isotropic and dispersive assuring that the TE$_{10}$ mode only propagates inside the waveguide. Port 1 launches an incident guided wave propagating along +z direction. The lateral dimensions of the waveguide are \textquotedblleft a\textquotedblright , \textquotedblleft b\textquotedblright~and the MUT fully fills the transverse cross section. The position of the short-circuit termination can be cyclically switched among a pre-determined set of locations $\left\{L_1, L_2, L_3, ...\right\}$. The repetition frequency ${{f}_{m}}={1}/{{{T}_{m}}}$ is far less than that of the propagating wave ${{f}_{0}}={1}/{{{T}_{0}}}$. This assumption allows us to use the adiabatic approximations \cite{Liu,Minkov}for putting the main foundation of our study on the same formulation already introduced for time-independent waveguide transmission lines.

Let us consider that the dielectric sample is illuminated by a monochromatic guided wave (dominant mode) with the transverse components of   

\begin{align}
& E_{y}^{i}(x,z)={{E}_{0}}sin({{\beta }_{x}}x){{e}^{-{{\gamma }_{z}}z}} \\ 
& H_{x}^{i}(x,z)={{E}_{0}}\frac{j{{\gamma }_{z}}}{\omega {{\mu }_{0}}}sin({{\beta }_{x}}x){{e}^{-{{\gamma }_{z}}z}}  
\end{align}
and suppressed time-harmonic dependence exp$(j2\pi f_0t)$. The time-domain reflected wave at the left side of the MUT can be written as 
\begin{align}
E_{y}^{r}(x,z,t)=E_{y}^{i}(x,t)\cdot S_{11}(t)
\end{align}
Since the position modulation of the short-circuit termination has a fixed period, the reflection coefficient can be mathematically described as a periodic function of time, and defined over one period as a linear combination of \textit{q} scaled and shifted pulses
\begin{align}
S_{11} \left( t \right)=\sum\limits_{n=1}^{q}{{{S_{11,n} }_{}}{{\Pi }_{n}}\left( t \right)~~~~~~~~0<t<T_m}
\end{align}
in which,
\begin{equation}
{{\Pi }_{n}}\left( t \right)= \left\{ \begin{array}{ll}
1      &~~\left( n-1 \right)\frac{{{T}_{m}}}{q}\le t\le n\frac{{{T}_{m}}}{q}\\
0      &~~~\text{o.w.}\end{array} \right.  
\end{equation}
and ${{S_{11,n}}_{}}$=${{r}_{n}}\measuredangle {{\phi }_{n}}$ denotes the complex reflection S-parameter, with ${r}_{n}$ and ${\phi}_{n}$ representing the reflection amplitude and phase during the \textit{n}$^{th}$ time interval (n$-$1)T$_m$/q$\le$\textit{t}$\le$nT$_m$/q, respectively. It should be pointed out that ${{r}_{n}}=1\left( \forall n\in \mathbb{N}+\{0\} \right)$ when loss-less samples are measured. Based on the impedance transformation concept \cite{Hasar2009}
\begin{align}
& Z_{in,1}^{n}=Z_{0}^{TE}\tanh ({{\gamma }_{z0}}{{L}_{n}}) \\ 
& Z_{in}^{n}=\frac{Z_{in,1}^{n}+Z_{d}^{TE}\tanh ({{\gamma }_{zd}}h)}{Z_{d}^{TE}+Z_{in,1}^{n}\tanh ({{\gamma }_{zd}}h)}\\
& {{S_{11,n}}_{}}=\frac{Z_{in}^{n}-Z_{0}^{TE}}{Z_{in}^{n}+Z_{0}^{TE}} 
\end{align}
Here, ${{\gamma }_{z0}}$=$\sqrt{{{\left( {\pi }/{a}\; \right)}^{2}}-k_{0}^{2}}$ and ${{\gamma }_{zd}}$=$\sqrt{{{\left( {\pi }/{a}\; \right)}^{2}}-k_{0}^{2}\epsilon_r \mu_r}$ indicate the z-directed wavenumbers inside the air-filled and sample-filled sections, respectively. Meanwhile, $L_n$ is the length of the air-filled section during the $n^{th}$ time interval, ${{k}_{0}}=2\pi f_0 \sqrt{{{\varepsilon }_{0}}{{\mu }_{0}}}$, $\varepsilon_0$, and $\mu_{0}$ refer to the wavenumber, permittivity, and permeability of the free-space, respectively, $\varepsilon_r$=$\varepsilon_r^\prime-j\varepsilon_r^{\prime\prime}$ and $\mu_r$=$\mu_r^\prime-j\mu_r^{\prime\prime}$ represent the relative complex permittivity and permeability of the sample, respectively, and finally, $Z_{0}^{TE}=j2\pi {{f}_{0}}{{\mu }_{0}}/{{\gamma }_{z0}}$ and $Z_{d}^{TE}=j2\pi {{f}_{0}}{{\mu }_{0}}/{{\gamma }_{zd}}$ stand for the characteristic impedances of the air-filled and sample-filled sections, respectively. The reference plane is arbitrarily set at the sample boundary (\textit{z}=0). As can be deduced from \textcolor{blue}{Eq. (6)}, a periodic change in the position of the short-circuit termination provides a straightforward route to dynamically modulate the equivalent impedance which circuitally loads the MUT \cite{Mirmoosa}. 
According to the periodic nature of the position modulation, $S_{11}(t)$ can be described through a Fourier series representation
\begin{align}
& S_{11} (t)=\sum\limits_{m=-\infty }^{\infty }{{{a}_{m}}{{e}^{j2\pi m{{f}_{m}}t}}}\\
& {{a}_{m}}=\sum\limits_{n=1}^{q}{\frac{S_{11,n} {{}}^{}}{m \pi} \text{sin}(\frac{m \pi}{q})\exp \left[ \frac{-jm \pi(2n-1)}{q} \right]}
\end{align}
The details of the derivation steps are summarized in \textcolor{blue}{Appendix}. The spectral information of the reflected field can be found by taking the Fourier transform of \textcolor{blue}{Eqs. (3), (9)} which yields
\begin{align}
& E_{y}^{r}\left( x,z,\omega  \right)=\frac{1}{2\pi }E_{y}^{i}\left( x,z,\omega  \right)*\Gamma \left( \omega  \right) \\
& S_{11} \left( \omega  \right)=\sum\limits_{m=-\infty }^{\infty }{{{a}^{m}}\delta \left( \omega -m{{\omega }_{m}} \right)} \\ 
& E_{y}^{r}\left( x,z,\omega  \right)=\frac{1}{2\pi }\sum\limits_{m=-\infty }^{\infty }{{{a}_{m}}\delta \left( \omega -{{\omega }_{0}}-m{{\omega }_{m}} \right)} 
\end{align}

Here, we assume a monochromatic excitation with the angular frequency $\omega_{0}$=$2 \pi f_0$. Following the temporal modulation, a nonlinear phenomenon occurs as characterized by the emergence of a series of harmonics around the central frequency, with phases/intensities ($a_m$) determined by the complex reflection S-parameter in each time slot \cite{Zhang, Zhao}. As an elucidating interpretation, the harmonic coefficients are a weighted average of the \textit{q} contributing terms with the information of the complex reflection S-parameter in each interval (\textcolor{blue}{Fig. 1}). The spectral location of harmonics, however, is solely dictated by the modulation frequency.

Now, we intend to focus on specific scenarios in which the position of the short-circuit termination is periodically switched between two distinct states, \textit{i.e.,} $\left\{L_1, L_2, L_1, L_2, ...\right\}$. In this case, $S_{11}$(t) becomes a square-wave signal with two different complex levels achieved via \textcolor{blue}{Eqs. (6)-(8)}. After some algebraic manipulations on \textcolor{blue}{Eq. (10)}, the harmonic coefficients read
\begin{align}
a_m= \left\{ \begin{array}{ll}
\frac{{r}_{1}}{2}{{\operatorname{e}}^{j{{\phi }_{1}}}}\left[ 1+\frac{{{r}_{2}}}{{{r}_{1}}}{{e}^{\Delta \phi }} \right]      &~~\text{synchronous component}\\
\frac{r_1}{jm\pi }{}{{\operatorname{e}}^{j{{\phi }_{1}}}}\left[ 1-\frac{{{r}_{2}}}{{{r}_{1}}}{{e}^{\Delta \phi }} \right]     &~~\text{odd harmonics}\\ 
0    &~~\text{even harmonics}\\ \end{array} \right. 
\end{align}
where, $\Delta \phi= \phi_1-\phi_2$ is the phase difference between the contributing reflection coefficients. Note that, arising from the Fourier properties of square waves, only the odd and zeroth harmonic components are found to survive in the reflected fields.

\subsection{Phase-Less Material Characterization}
Before presenting the proposed method, a conventional time-independent configuration for electromagnetic parameter retrieval in short-circuited waveguides is schematically depicted in \textcolor{blue}{Fig. 2}. In this method which uses both amplitude and phase data, the unknown parameters are the frequency spectra of  $\varepsilon_r$=$\varepsilon_r^\prime-j\varepsilon_r^{\prime\prime}$ and $\mu_r$=$\mu_r^\prime-j\mu_r^{\prime\prime}$ while the known data are the amplitude and phase spectra of the reflection S-parameter upon two different positions of the short-circuit termination (S$_{11,1}$ and S$_{11,2}$), which are measured at the reference-plane. Such a material characterization method usually deals with two major problems. First, when the sample is medium- or low-loss, electromagnetic waves will attenuate less while passing through it, and hence they will bounce back and forth inside the sample (multiple reflections) \cite{Hasar2010}. In this circumstance, it is difficult to distinguish the reflected signals from the sample front surface with those transmitted through the same surface after re-reflecting inside the sample, and thus uneasy to determine unique MUT parameters. Here, to ensure achieving unique solution without need for a priori information, all defined measurements are performed for two samples of the same MUT (identical homogeneity and electromagnetic parameters) with different thicknesses (\textcolor{blue}{Fig. 3}). This requirement can be easily provided through cutting the prepared sample. If the MUT is a lossy sample (lossy enough producing at least 10 dB attenuation), the effect of transcendental terms can be automatically eliminated and the second measurement is no longer required. Also, this auxiliary test is not necessary, provided that we have a good initial guess about the range of $\varepsilon_r$ and/or $\mu_r$. The uniqueness analysis has been comprehensively reported in \cite{Hasar2017}. As the second problem, the direct phase measurement is a difficult procedure to carry out, as, in addition to high cost and heavy time consumption, it generally requires that the reference-plane of the MUT be accurately defined. Any uncertainty in the calibration-plane leads to considerable errors, especially at high frequencies and for measurements accomplished far away from the measurement-plane. 
 
  \begin{figure}[t]
 	\includegraphics[scale=0.32]{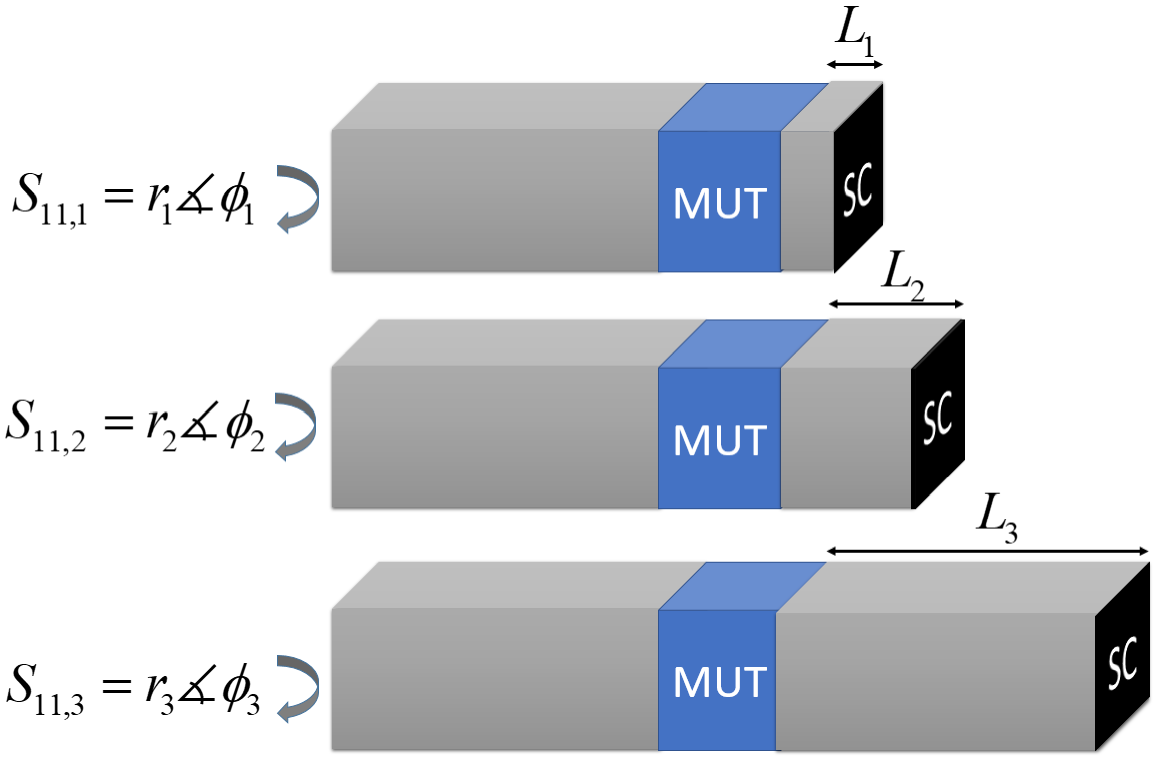}
 	\caption{\label{fig:epsart} {Flowchart of the reconstruction algorithm  by using amplitude-only reflection measurements of a time-modulated waveguide.} }
 \end{figure}

In this paper, as an alternative approach, we will exploit only the amplitude information of the harmonics ($|a_m|$) emerging in the proposed time-modulated waveguide configuration to indirectly extract the accurate phase of the reflection S-parameter, thereby circumventing any difficulties and uncertainties accompanied with the phase calibration procedures. Our idea is graphically sketched in \textcolor{blue}{Fig. 1}. Referring to \textcolor{blue}{Eq. (14)}, the harmonic intensities resulted by a dual-state temporal modulation can be written as
\begin{align}
|a_m|= \left\{ \begin{array}{ll}
\frac{{r}_{1}}{2}\left| 1+\frac{{{r}_{2}}}{{{r}_{1}}}{{e}^{\Delta \phi }} \right|       &~~\text{synchronous component}\\
\frac{{{r}_{1}}}{\left| m \right|\pi }\left| 1-\frac{{{r}_{2}}}{{{r}_{1}}}{{e}^{\Delta \phi }} \right|     &~~\text{odd harmonics}\\ 
0    &~~\text{even harmonics}\\ \end{array} \right. 
\end{align}
In the case of loss-less (or extremely low-loss) MUT, $r_1$$\simeq$$r_2$$\simeq$$ 1$, and the above expressions can be reduced to 
\begin{align}
|a_m|_{\text{LL}}= \left\{ \begin{array}{ll}
\left| \cos (\frac{\Delta \phi }{2}) \right|     &~~\text{synchronous component}\\
\frac{2}{\left| m \right|\pi }\left| \sin (\frac{\Delta \phi }{2}) \right| &~~\text{odd harmonics}\\ 
0    &~~\text{even harmonics}\\ \end{array} \right. 
\end{align}

From \textcolor{blue}{Eqs. (15), (16)}, the following remarks can elaborately be concluded:

1) The harmonic intensities are a nonlinear function of phase difference $\Delta \phi$, not the absolute phases $\phi_{1}$ and $\phi_{2}$. Therefore, the phase information of $S_{11,1}$ and $S_{11,2}$ can not be recovered, separately, and the number of equations is decreased by one.
 
2) Since the intensities of odd harmonic frequencies differ just by the harmonic order factor, one of them can be employed as the independent information, only.

3) Given trigonometry formulas, the amplitude information of the synchronous component and odd harmonics are not independent in the case of loss-less (or extremely low-loss) dielectric samples.

4) The parameters $r_1$ and $r_2$ can be simply measured in a phase-less scheme when the short-circuit termination is statically located at $L_1$ and $L_2$, respectively. 

5) The profile of harmonic intensities are an even function of harmonic order (m) meaning that the positive and negative harmonics do not reveal independent data. 

In overall, $\Delta \phi$ is the key unknown parameter in the inverse problem which manifests itself into the measured harmonic intensities. According to the first remark, the reduction in number of equations can be compensated by appending a new test configuration in which the position of short-circuit termination is cyclically swapped between $L_1$ and $L_3$ with modulation frequency $f_m$. Consequently,  $\Delta \phi_{12}$=$\phi_{1}-\phi_{2}$ and $\Delta \phi_{13}$=$\phi_{1}-\phi_{3}$ should be inevitably taken as the unknown parameters in the inverse problem instead of $\phi_1$ and $\phi_2$. After delving into the nonlinear nature of \textcolor{blue}{Eqs. (15), (16)}, one can deduce that two possible solutions with opposite signs may be obtained for the phase difference parameters, especially in the case of loss-less samples for which only a single harmonic in each of \textcolor{blue}{Eqs. (15), (16)} reveals instrumental information. Thus, an additional measurement with triple-state switching must be planned to identify the correct sign of $\Delta \phi_{12}$ and $\Delta \phi_{13}$, wherein the sequence $\left\{L_1, L_2, L_3, L_1, L_2, L_3, ...\right\}$ indicates the position of short-circuit termination over the temporal dimension. Care should be taken that this time sequence is adopted to ensure the uniqueness of $\Delta \phi_{ij}$ extraction while the thickness diversity is applied to guarantee the uniqueness of reconstruction of the electromagnetic parameters. The harmonic intensities in such a configuration obey \textcolor{blue}{Eq. (17)}. As can be seen, the remarks 2, 3, and 5 are no longer valid in the case of triple-state switching. We should also remember that the amplitude factors $r_1$, $r_2$, and $r_3$ are already measured when the short-circuit termination is placed at $L_1$, $L_2$, and $L_3$ positions, respectively, without any temporal change.

\begin{widetext}
	\begin{align}
	|a_m|= \left\{ \begin{array}{ll}
	\frac{{{r}_{1}}}{3}\left| 1+\frac{{{r}_{2}}}{{{r}_{1}}}{{e}^{-j\Delta {{\phi }_{12}}}}+\frac{{{r}_{3}}}{{{r}_{1}}}{{e}^{-j\Delta {{\phi }_{13}}}} \right|  &~~\text{synchronous component}\\
	\frac{{{r}_{1}}}{\pi }\left| \frac{1}{m}\sin (\frac{m\pi }{3}) \right|\left| 1+\frac{{{r}_{2}}}{{{r}_{1}}}{{e}^{-j\left( \Delta {{\phi }_{12}}+\frac{2m\pi }{3} \right)}}+{{\frac{{{r}_{3}}}{{{r}_{1}}}}^{-j\left( \Delta {{\phi }_{13}}+\frac{4m\pi }{3} \right)}} \right|   &~~m\ne \pm 3,\,\pm 6,\,\,\pm 9,...\\ 
	0  &~~m=\pm 3,\,\pm 6,\,\,\pm 9,...\\ \end{array} \right. 
	\end{align}
\end{widetext}

To sum up, the flowchart for the proposed scalar material characterization using the time-modulated waveguides is shown in \textcolor{blue}{Fig. 3}. In order to extract the phase difference factors $\Delta \phi_{12}$ and $\Delta \phi_{13}$, uniquely, three distinct measurements are adopted in this paper, wherein $\left\{L_1, L_2, L_1, L_2, ...\right\}$, $\left\{L_1, L_3, L_1, L_3, ...\right\}$, and $\left\{L_1, L_2, L_3, L_1, L_2, L_3, ...\right\}$ are the time program for the position modulation of the short-circuit termination. The magnitudes of reflection S-parameters $r_1$, $r_2$, and $r_3$ are initially simulated or measured. In each measurement, the intensity of one of the non-trivial harmonics is recorded as independent data. The harmonic order can be arbitrarily chosen by considering different criteria such as frequency isolation, ease of recording, strength, and so on. As emphasized, these measurements are repeated for two identical samples with different thicknesses $h_1$ and $h_2$ to yield the unknown parameters $\Delta \phi_{12,h_1}$, $\Delta \phi_{13,h_1}$, $\Delta \phi_{12,h_2}$, and $\Delta \phi_{13,h_2}$. The systems of nonlinear equations for two thicknesses are quite independent and can be solved, separately. Indeed, for each thickness, we have a system of three coupled equations to extract $\Delta \phi_{12,h_i}$ and $\Delta \phi_{13,h_i}$ ($i=0,1$). A fast optimization is adopted to numerically solve both systems of equations through minimizing the error between measured/simulated and analytically calculated harmonic intensities (\textcolor{blue}{Eq. (18)}). The subscript \textquotedblleft TH\textquotedblright~remarks the theoretical values governed by \textcolor{blue}{Eqs. (15), (17)}, while the subscript \textquotedblleft SIM\textquotedblright~denotes the simulated or measured harmonic intensities. The lossy materials can be supported by more number of independent equations provided by different harmonic orders to be robust against noise or any systematic and/or random errors. In this case, a system of six coupled equations (two of harmonic intensities in each measurement) is provided for each thickness.

\begin{widetext}
	\begin{align}
	E_1=\left( {{\left| {{a}_{i}} \right|}_{\text{TH}}}-{{\left| {{a}_{i}} \right|}_{\text{SIM}}} \right)_{{{L}_{1}}{{L}_{2}}...}^{2}+\left( {{\left| {{a}_{j\ne i}} \right|}_{\text{TH}}}-{{\left| {{a}_{j\ne i}} \right|}_{\text{SIM}}} \right)_{{{L}_{1}}{{L}_{2}}...}^{2}+\left( {{\left| {{a}_{i}} \right|}_{\text{TH}}}-{{\left| {{a}_{i}} \right|}_{\text{SIM}}} \right)_{{{L}_{1}}{{L}_{3}}...}^{2} \\ 
	\nonumber+\left( {{\left| {{a}_{j\ne i}} \right|}_{\text{TH}}}-{{\left| {{a}_{j\ne i}} \right|}_{\text{SIM}}} \right)_{{{L}_{1}}{{L}_{3}}...}^{2}+\left( {{\left| {{a}_{i}} \right|}_{\text{TH}}}-{{\left| {{a}_{i}} \right|}_{\text{SIM}}} \right)_{{{L}_{1}}{{L}_{2}}{{L}_{3}}...}^{2}+\left( {{\left| {{a}_{j\ne i}} \right|}_{\text{TH}}}-{{\left| {{a}_{j\ne i}} \right|}_{\text{SIM}}} \right)_{{{L}_{1}}{{L}_{2}}{{L}_{3}}...}^{2}  
	\end{align}
\end{widetext}

After obtaining $\Delta \phi_{12,h_1}$, $\Delta \phi_{13,h_1}$, $\Delta \phi_{12,h_2}$, and $\Delta \phi_{13,h_2}$ from the simulated or measured data, \textcolor{blue}{Eqs. (6)-(8)} are established to construct the required system of equations for determination of $\varepsilon_r$=$\varepsilon_r^\prime-j\varepsilon_r^{\prime\prime}$ and $\mu_r$=$\mu_r^\prime-j\mu_r^{\prime\prime}$. In the case of lossy materials, the magnitudes of reflection S-parameters, $r_1$, $r_2$, and $r_3$ also add three independent equations. Thereupon, we have a system of seven nonlinear equations with four unknown parameters $\varepsilon_r^\prime$, $\varepsilon_r^{\prime\prime}$, $\mu_r^\prime$, and $\mu_r^{\prime\prime}$ which can be numerically solved by minimizing the cost function of \textcolor{blue}{Eq. (19)}. The superscript \textquotedblleft TH\textquotedblright~ refers to the theoretical values governed by \textcolor{blue}{Eqs. (6)-(8)}, while the superscript \textquotedblleft EXT\textquotedblright~ represents the values extracted from solving \textcolor{blue}{Eq. (8)}. The detailed expressions for $r^{\text{TH}}_n$ and $\Delta \phi^{\text{TH}}_{ij}$ are given in \textcolor{blue}{Appendix}. Moreover, $w_p$ and $w_a$ represent the weight coefficients that can be established to selectively fine the phase and amplitude terms, respectively. Note that the whole process can be simply repeated for each microwave frequency to afford the spectral information of the unknown electromagnetic properties which are of great importance for highly-dispersive dielectric samples.

\begin{widetext}
	\begin{align}
	{{E}_{2}}=w_p{{\left( \Delta \phi _{12,{{h}_{1}}}^{\text{TH}}-\Delta \phi _{12,{{h}_{1}}}^{\text{EXT}} \right)}^{2}}+w_p{{\left( \Delta \phi _{13,{{h}_{1}}}^{\text{TH}}-\Delta \phi _{13,{{h}_{1}}}^{\text{EXT}} \right)}^{2}}+w_p{{\left( \Delta \phi _{12,{{h}_{2}}}^{\text{TH}}-\Delta \phi _{12,{{h}_{2}}}^{\text{EXT}} \right)}^{2}}+w_p{{\left( \Delta \phi _{13,{{h}_{2}}}^{\text{TH}}-\Delta \phi _{13,{{h}_{2}}}^{\text{EXT}} \right)}^{2}}\\
	\nonumber+w_r{{\left( r_{1}^{\text{TH}}-r_{1}^{\text{SIM}} \right)}^{2}}+w_r{{\left( r_{2}^{\text{TH}}-r_{2}^{\text{SIM}} \right)}^{2}}+w_r{{\left( r_{3}^{\text{TH}}-r_{3}^{\text{SIM}} \right)}^{2}}
	\end{align}
\end{widetext}

The numerical technique \textquoteleft fmincon\textquoteright~function in MATLAB is applied to \textcolor{blue}{Eqs. (18), (19)}, drawing out the electromagnetic parameters of the MUT with a fast convergence rate and without resorting to an initial guess. This function is applicable for nonlinear expressions while allowing some restrictions on the unknown quantities. 
\begin{figure}[h]
	\includegraphics[scale=0.23]{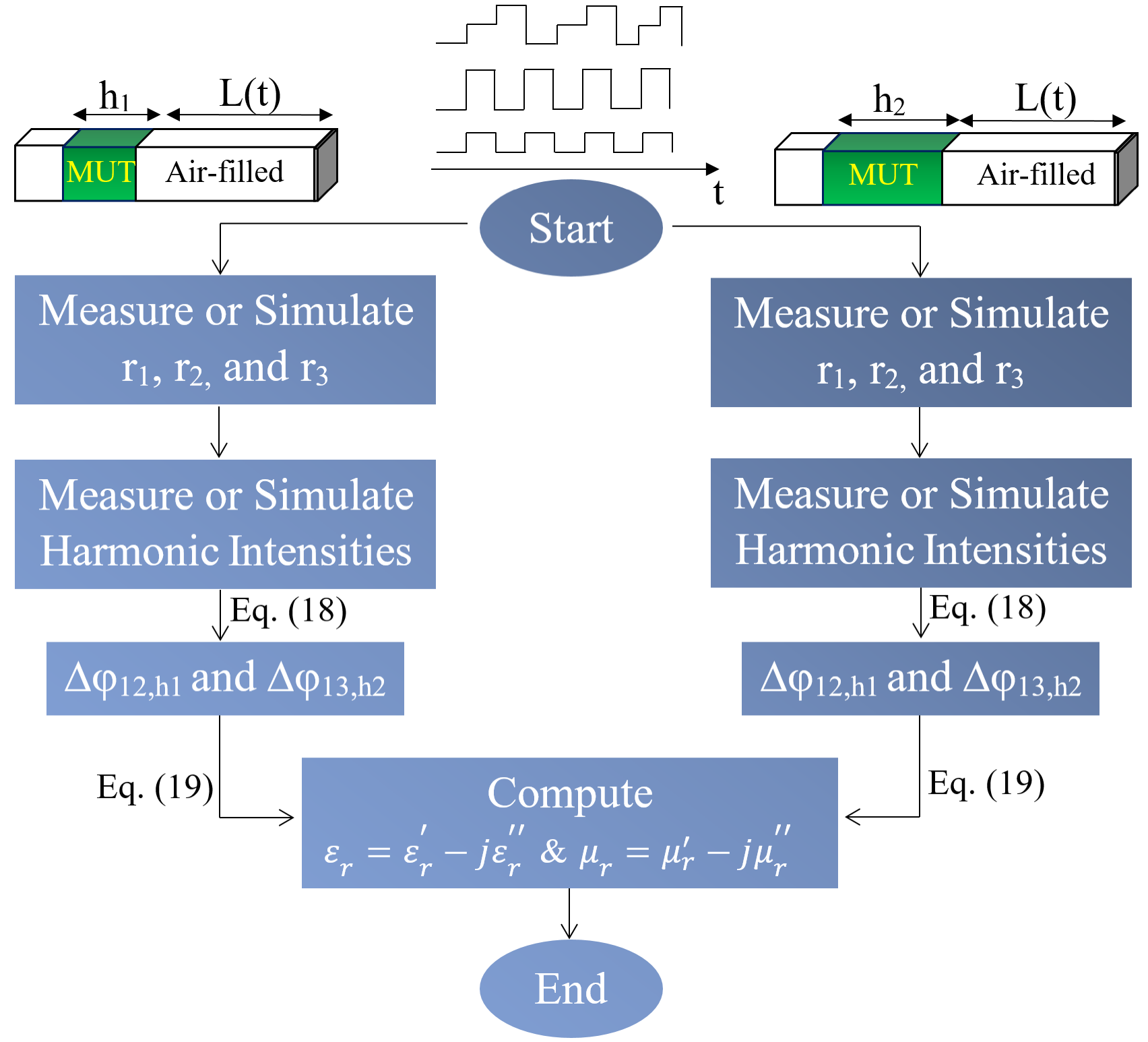}
	\caption{\label{fig:epsart} {Flowchart of the proposed reconstruction algorithm  by using amplitude-only reflection measurements of a time-modulated waveguide.} }
\end{figure}

\section{Results and Discussion}
In this section, the validity of the devised method is investigated through different illustrative examples. In our demonstrations, we used a WR-90 waveguide cell with length of 44.38 mm and cross section of 22.86$\times$10.16 mm$^2$ to determine the electromagnetic parameters using temporal modulation. For each example, the simulated data are computed in several equally distributed distinct points from $f_0$=8 to 12 GHz. To provide independent data with acceptable discrimination, the position of the short-circuit termination is temporally modulated between $L_1$=0 mm, $L_2$=5 mm, and $L_3$=10 mm. It is worth mentioning that this set of positions can be adaptively changed for each frequency to increase the accuracy of the results, especially for high refractive index or very lossy materials. The magnitudes of the reflection S-parameter, $r_1$, $r_2$, and $r_3$, are numerically calculated by using a waveguide measurement setup built in a commercial software, CST Microwave Studio (\textcolor{blue}{Fig. 4}). The corresponding results are then fed into our retrieval algorithm. The frequency gap between the adjacent harmonics should be large enough to be detected by the commercial spectrum analyzers. A comprehensive discussion will be presented in Section IV. The constraints we applied in the computation were $-\pi$$<$$\Delta \phi_{ij}$$<$$\pi$, $1$$<$$\varepsilon^{\prime}_r$$<$$20$, $1$$<$$\mu^{\prime}_r$$<$$20$, $0$$<$$\tan {{\delta }_{e}}={{\varepsilon^{\prime \prime}_r}}/{{\varepsilon^{\prime}_r}}$$<$$1$, and $0$$<$$\tan {{\delta }_{m}}={{\mu^{\prime \prime}_r}}/{{\mu^{\prime}_r}}$$<$$1$. The algorithm we applied was the Levenberg-Marquardt with maximum iterations 1000, function tolerance 10$^6$, and maximum function evaluations 10$^8$. The phase and amplitude weight factors are selected as $w_p$=${1}/{2\pi }\;$ and $w_a$=2, respectively. The unique solutions can be conveniently extracted from the solution domain by using our phase-less formalism especially for materials whose electromagnetic properties are not known a priori.
\subsection{Example I: Non-magnetic Lossy Materials}
In the first example, different non-magnetic  practical materials listed in \textcolor{blue}{Table. I} are considered to be characterized with two unknown parameters ($\varepsilon_r$ and tan$\delta_e$). To inspect the validity range of the proposed algorithm, a broad range of permittivity $\in$$ \left[ 2,20 \right]$ and loss tangent $\in$$ \left[ 0,1 \right]$ quantities are surveyed with diverse thicknesses. The harmonic components captured in the reflection side of the short-circuited waveguide are illustrated in \textcolor{blue}{Figs. 5a-c}, corresponding to the position sequences of $\left\{L_1, L_2, L_1, L_2, ...\right\}$, $\left\{L_1, L_3, L_1, L_3, ...\right\}$, and $\left\{L_1, L_2, L_3, L_1, L_2, L_3, ...\right\}$, respectively. The results are given for a monochromatic incidence at $f_0$=10 GHz. For the sake of brevity, the results are displayed only for Arlon CLTE-AT of thickness $h_1$=3 mm when excited by a monochromatic guided wave at 10 GHz. One of harmonic intensities in each figure along with the simulated magnitudes of reflection S-parameter are the input data for solving \textcolor{blue}{Eq. (18)}. Retrieved spectra of $\Delta \phi_{13,h_1}$ and $\Delta \phi_{12,h_1}$ are depicted in \textcolor{blue}{Fig. 5d}. For the sake of comparison, the exact phase differences are numerically recorded by the configuration of \textcolor{blue}{Fig. 4}. Excellent agreement between the results confirms that, by using the proposed time-modulated scheme, the relative phases of the reflection S-parameter are perfectly elicitable from the harmonic intensities. The inverse problem is proceed with feeding the outcomes into \textcolor{blue}{Eq. (19)} to reconstruct the spectral information of $\varepsilon_r$ and $\tan \delta_e$. The synthetic harmonic intensities corresponding to different specimens of \textcolor{blue}{Table. I} are numerically generated to extract their dielectric constant using the presented algorithm. We assume the auxiliary tests are accomplished for the same MUT with thickness $t_2$ given in \textcolor{blue}{Table. I}. The extracted $\varepsilon_r$ and $\tan \delta_e$ for all these reference samples are given in \textcolor{blue}{Figs. 6a-c}, respectively. Our demonstrations are performed for different ranges of $\varepsilon_r$, $\tan \delta_e$ and $h$. As noticed, the retrieved parameters for all considered samples are quite close to their original data. This observation declares that the proposed approach relying on the temporal modulation successfully serve to procure the broadband complex permittivity of specimens from simple amplitude-only measurements followed by a fast post-processing step.

\begin{figure}[h]
	\includegraphics[scale=0.3]{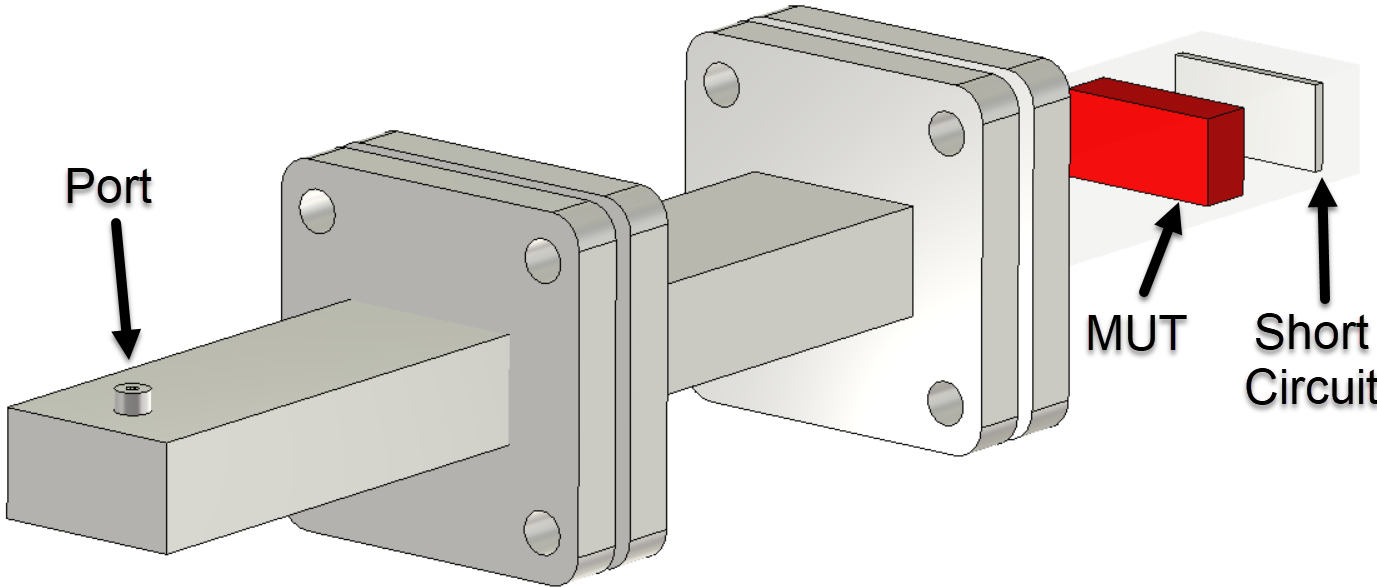}
	\caption{\label{fig:epsart} {The simulation or measurement waveguide setup to extract the amplitude of reflection S-parameter. } }
\end{figure}

\begin{table}[h]
	\caption{Dielectric properties of several lossy and non-magnetic materials which are characterized to verify the validity of the proposed reconstruction algorithm.  }
	\label{tbl:example}
	\centering
	\begin{tabular}{ c|c|c|c|c|c  }
		\multicolumn{4}{c}{} \\ 
		MUT& $\varepsilon_r$ & $\tan \delta_e$ & h$_1$ (mm)& h$_2$ (mm)&Type \\ 
		\hline
		PTFE& 2.1&0.0002& 6& 2& A1 \\
		\hline
		Arlon CLTE-AT& 3 & 0.0013& 1.5& 0.5& A2 \\
		\hline
		Arlon AD410& 4.1 & 0.003& 3& 0.8& A3 \\
		\hline
		Preperm L500& 5 & 0.0005& 5& 2& A4 \\
		\hline
		Taconic RF-60A& 6.15 & 0.0028& 1.6& 0.6& A5\\
		\hline
		Preperm L700HF& 7 & 0.0006& 4.5& 3& A6 \\
		\hline
		Aluminum Nitride& 8.6  & 0.0003& 2& 0.8& A7\\
		\hline
		Taconic CER-10& 10 & 0.0035 & 3.2& 0.75& A8\\
		\hline
		Rogers RO3010& 11.2  & 0.0022& 1.3& 0.25& A9\\
		\hline
		Rogers TMM 13i& 12.2 & 0.0019 & 3.8& 2.5& A10\\
		\hline
		Sandy Soil (18.8\% water)& 13&0.29& 6& 5& A11 \\
		\hline
		Loamy Soil (13.77\% water)& 13.8&0.18& 4& 3& A12 \\
		\hline
		Sample 1& 15& 0.45& 4.5& 3& A13 \\
		\hline
		Sample 2& 16.5& 0.6& 5.5& 3.5& A14 \\
		\hline
		Sample 3& 18& 0.75& 2.5& 1.5& A15 \\
		\hline
		Sample 4& 19.5&0.9&3.5 & 2& A16 \\
		\hline
	\end{tabular}
\end{table}

\subsection{Example II: Electromagnetic Lossy Dispersive Materials}
Up to now, the presented demonstrations have been devoted to non-magnetic materials for which the constitutive parameters do not vary over the frequency band of study, significantly. Here, we intend to assess the proposed algorithm for a more intricate category of materials with four frequency-dependent unknown parameters, \textit{i.e.,} electromagnetic lossy dispersive materials. In many real-life applications, for instance PCB substrates \cite{Zhangieee}, radar absorbing composites \cite{rahman}, and metamaterials \cite{smith}, it is necessary to know the frequency-dependent behavior of the bulk material constituting the complex structures. The well-known dispersion laws obeying the Kramers-Kronig causality relations, \textit{e.g.,} Debye or Lorentzian model, can be used to fit the spectral trend of the permittivity and permeability parameters.

\begin{figure*}[t]
	\includegraphics[scale=0.3]{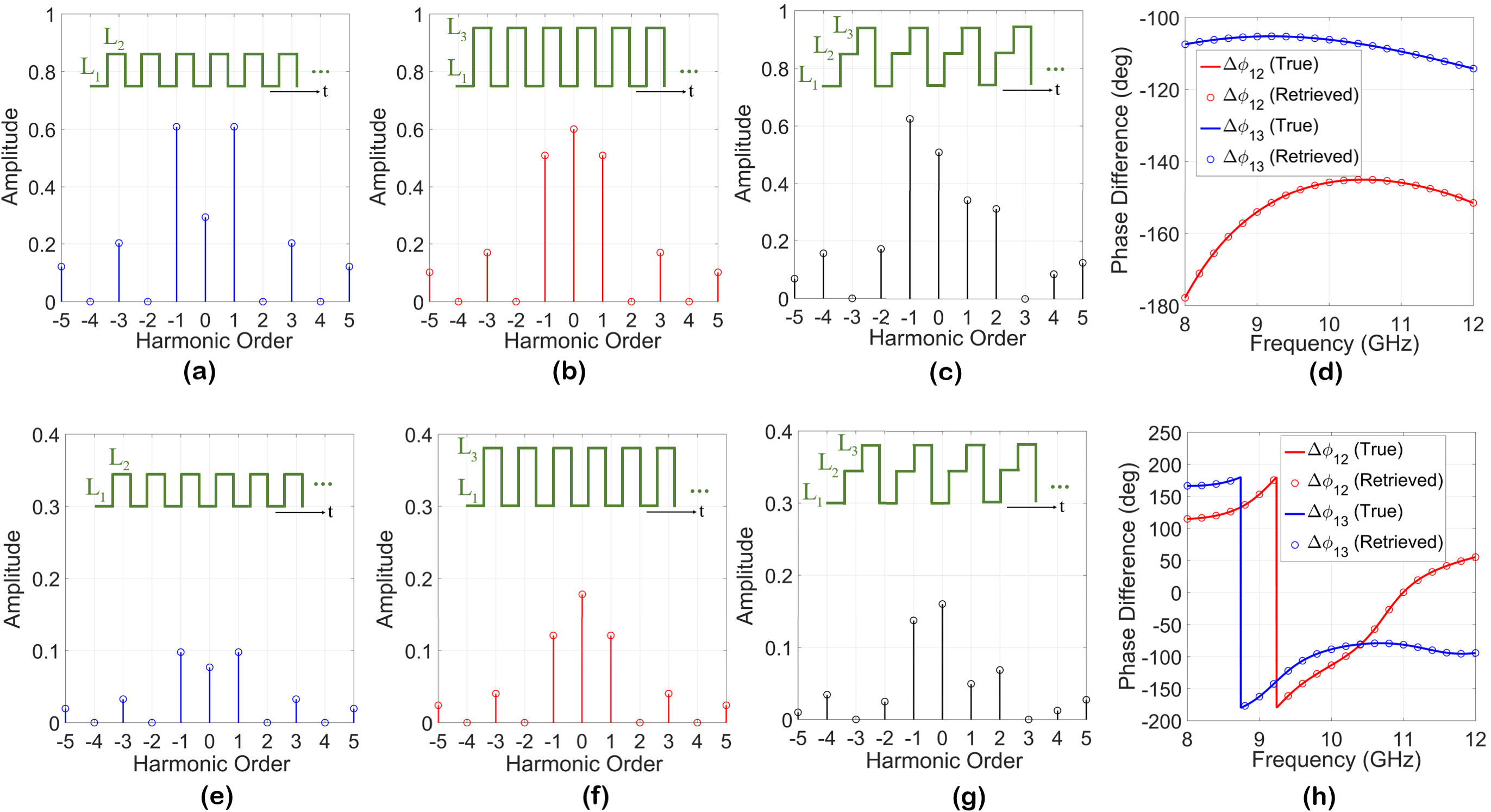}
	\caption{\label{fig:epsart} {The recorded amplitudes of (a)-(c), (e)-(g) reflected harmonics, when the position of short-circuit termination is modulated over the temporal dimension with the sequences of (a), (e) $\left\{L_1, L_2, L_1, L_2, ...\right\}$, (c), (g) $\left\{L_1, L_3, L_1, L_3, ...\right\}$, and (b), (f) $\left\{L_1, L_2, L_3, L_1, L_2, L_3,...\right\}$}. (d), (h) A comparison between the true and retrieved phase differences of $\Delta\phi_{12}$ and $\Delta\phi_{13}$, where the results (a)-(d) and (e)-(h) correspond to samples A2 and B2 of thickness $h_1$, respectively. The harmonic magnitudes are plotted for a monochromatic incidence at $f_0$= 10 GHz.}
\end{figure*}

The real and imaginary parts of the electromagnetic parameters for a linear, isotropic, and homogeneous dispersive material can be mathematically expressed by a one-term Debye model \cite{Zhangieee}
\begin{align}
& {\varepsilon }'\left( \omega  \right)=\frac{{{\varepsilon }_{s}}+{{\varepsilon }_{\infty }}{{\left( \omega \tau_e  \right)}^{2}}}{1+{{\left( \omega \tau_e  \right)}^{2}}}\\
& {\varepsilon }''\left( \omega  \right)=\frac{\left( {{\varepsilon }_{s}}-{{\varepsilon }_{\infty }} \right)\omega \tau_e }{1+{{\left( \omega \tau_e  \right)}^{2}}}+\frac{{{\sigma }_{e}}}{\omega {{\varepsilon }_{0}}} \\ 
& {\mu }'\left( \omega  \right)=\frac{{{\mu }_{s}}+{{\mu }_{\infty }}{{\left( \omega \tau_m  \right)}^{2}}}{1+{{\left( \omega \tau_m  \right)}^{2}}}\\
& {\mu }''\left( \omega  \right)=\frac{\left( {{\mu }_{s}}-{{\mu }_{\infty }} \right)\omega \tau_m }{1+{{\left( \omega \tau_m  \right)}^{2}}} 
\end{align} 

or a single-component Debye form 

\begin{align}
& {{{\varepsilon }'}}\left( \omega  \right)=\frac{\left( \omega _{0,e}^{2}-{{\omega }^{2}} \right)\left( {{\varepsilon }_{s}}\omega _{0,e}^{2}-{{\varepsilon }_{\infty }}{{\omega }^{2}} \right)+4{{\varepsilon }_{\infty }}{{\left( \omega \delta_e  \right)}^{2}}}{{{\left( \omega _{0,e}^{2}-{{\omega }^{2}} \right)}^{2}}+4{{\left( \omega \delta_e  \right)}^{2}}}\\
& {{{\varepsilon }''}}\left( \omega  \right)=\frac{2\omega \delta_e \omega _{0,e}^{2}\left( {{\varepsilon }_{s}}-{{\varepsilon }_{\infty }} \right)}{{{\left( \omega _{0,e}^{2}-{{\omega }^{2}} \right)}^{2}}+4{{\left( \omega \delta_e  \right)}^{2}}}+\frac{{{\sigma }_{e}}}{\omega {{\varepsilon }_{0}}} \\ 
& {{{\mu }'}}\left( \omega  \right)=\frac{\left( \omega _{0,m}^{2}-{{\omega }^{2}} \right)\left( {{\mu }_{s}}\omega _{0,m}^{2}-{{\mu }_{\infty }}{{\omega }^{2}} \right)+4{{\mu }_{\infty }}{{\left( \omega \delta_m  \right)}^{2}}}{{{\left( \omega _{0,e}^{2}-{{\omega }^{2}} \right)}^{2}}+4{{\left( \omega \delta_m  \right)}^{2}}}\\
& {{{\mu }''}}\left( \omega  \right)=\frac{2\omega \delta_m \omega _{0,m}^{2}\left( {{\mu }_{s}}-{{\mu }_{\infty }} \right)}{{{\left( \omega _{0,m}^{2}-{{\omega }^{2}} \right)}^{2}}+4{{\left( \omega \delta_m  \right)}^{2}}}
\end{align}

\begin{figure*}[t]
	\includegraphics[scale=0.29]{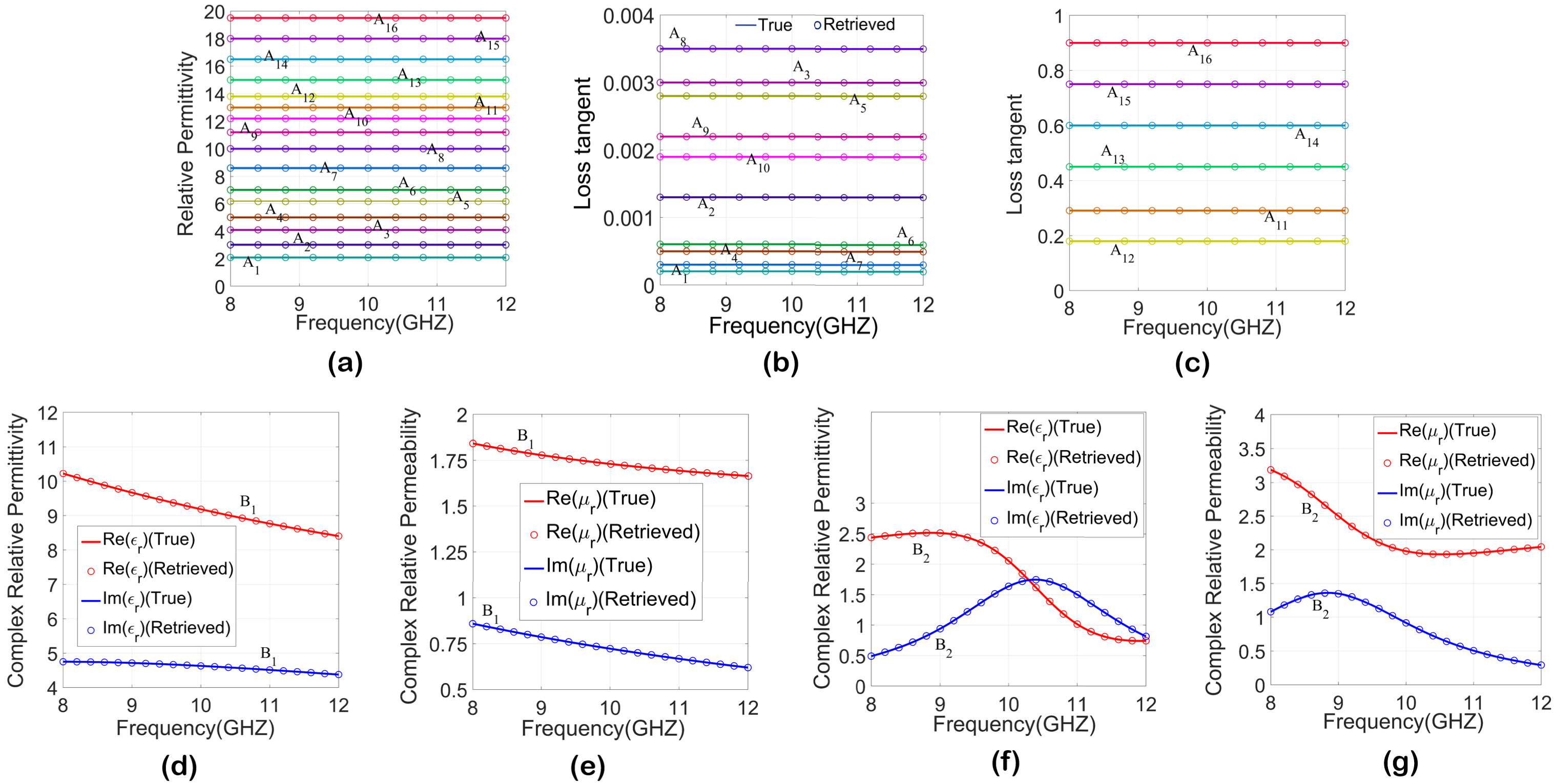}
	\caption{\label{fig:epsart} {The exact and retrieved values of (a) the relative permittivity and (b), (c) loss tangent of 16 different lossy and non-magnetic MUTs listed in Table. I. The extract and retrieved values of real and imaginary parts of (d), (f) relative permittivity and (e), (g) permeability dispersive samples listed in Table. II, which are characterized by (d), (e) Debye and (f), (g) Lorentzian dispersion model. } }
\end{figure*}

Here, the subscripts \textquotedblleft s\textquotedblright~and \textquotedblleft $\infty$\textquotedblright~represent the relative electromagnetic parameters at the static and high frequencies, respectively, $\tau_{e/m}$ is the relaxation time, $\omega_{0,e/m}$=2$\pi$$f_{0,e/m}$ and $\delta_{e/m}$ indicate the electric/magnetic resonant frequency and the width of Lorentzian peak, respectively, and finally, $\sigma_e$ denotes the electrical conductivity of MUT. We consider two isotropic lossy specimens with $h_2$=$h_1$/3=1 mm and different dispersion models specified in \textcolor{blue}{Table. II}. The same procedure discussed in the previous section is sequentially carried out to extract $\varepsilon_r\left(\omega\right)$=$\varepsilon_r^\prime\left(\omega\right)-j\varepsilon_r^{\prime\prime}\left(\omega\right)$ and $\mu_r\left(\omega\right)$=$\mu_r^\prime\left(\omega\right)-j\mu_r^{\prime\prime}\left(\omega\right)$ except that the amplitude levels of two harmonics including the synchronous component are involved in each measurement. Starting from $L$=0 mm, the step size of automatic position variation for the movable short-circuit termination is set as 5 mm. \textcolor{blue}{Figs. 5e-g} illustrate the amplitudes of the harmonic components as the input data of inverse problem. The magnitudes of the reflection S-parameter are numerically recorded by using the waveguide setup shown in \textcolor{blue}{Fig. 4}. Following the flowchart of \textcolor{blue}{Fig. 3}, the calculated values of $\Delta \phi_{12,h_1}$ and $\Delta \phi_{13,h_1}$ are compared with the exact solutions in \textcolor{blue}{Fig. 5h}. The true and reconstructed values of the real and imaginary parts of the permittivity and permeability spectra are also demonstrated in \textcolor{blue}{Figs. 6d-g}. The retrieved data follow the spectral trend of the benchmark parameters, perfectly, revealing that the proposed time-modulated material characterization approach can be successfully applied to highly dispersive electromagnetic samples through executing amplitude-only measurements. The uniqueness of solutions have been examined for a set of 5000 distinct electromagnetic parameters for each of which, the computational time is about 4s (RAM 16.0 GB and CPU Intel(R) i7-6700 HQ $@$2.60 GHz). We should emphasize that using the suggested retrieval scheme, there is no necessity to know a priori information regarding the electromagnetic parameters of MUT.

\begin{figure*}[t]
	\includegraphics[scale=0.42]{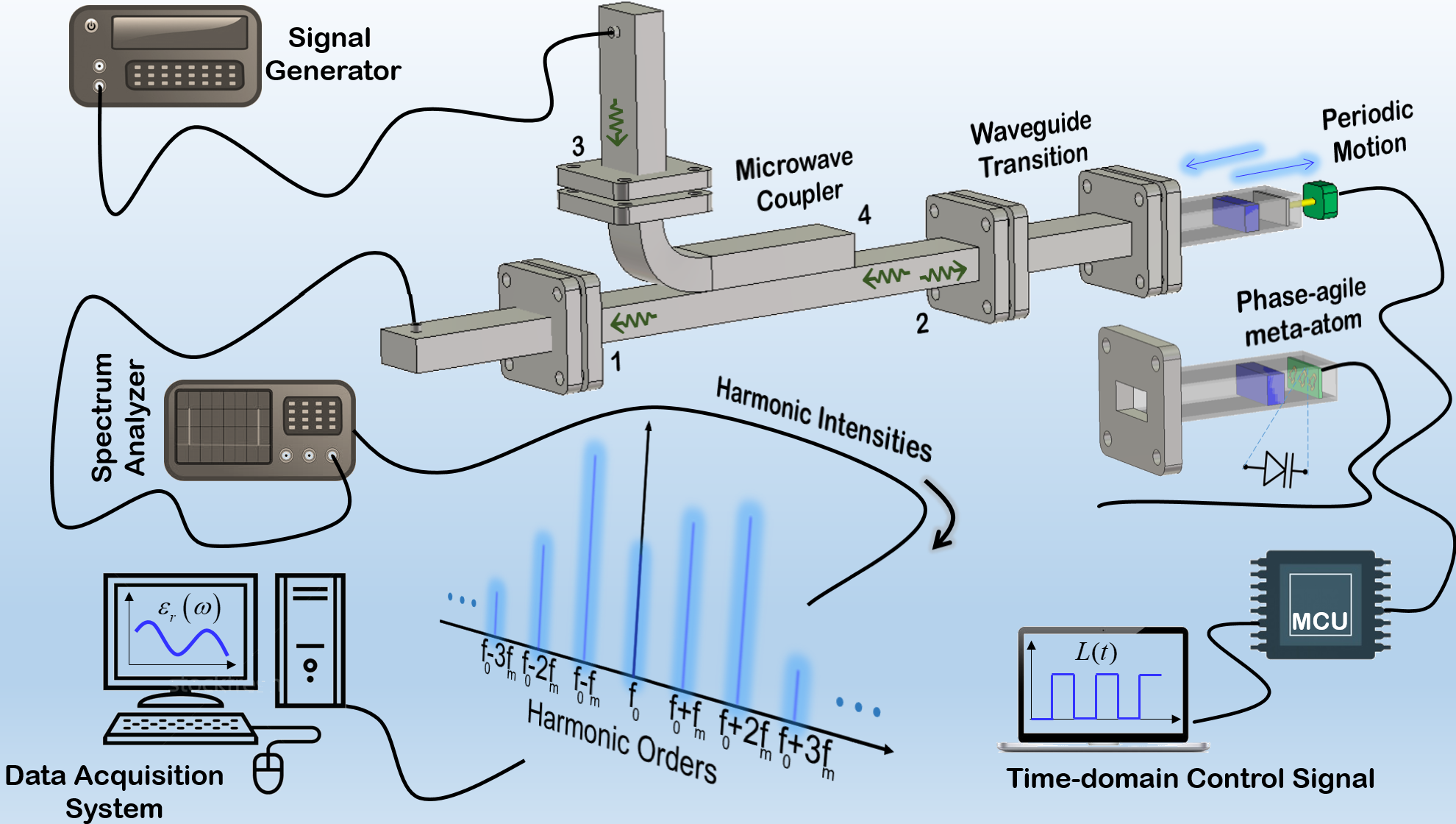}
	\caption{\label{fig:epsart} {Two possible realization schemes of the time-modulated waveguides to execute the proposed electromagnetic parameter retrieval method using amplitude-only reflection measurements. The signal generator excites the MUT via a monochromatic illuminating wave while the short-circuit termination is periodically modulated through a time-varying sequence generated by the MCU. The intensity of emerging harmonics are recorded by means of the coupler and spectrum analyzer. The electromagnetic parameters are reconstructed after the measured data being processed by the data acquisition system. } }
\end{figure*}

\begin{table}[h]
	\caption{Electromagnetic parameters of two lossy MUTs characterized by Debye and Lorentzian dispersion forms.}
	\label{tbl:example}
	\centering
	\begin{tabular}{ c|c|c|c|c|c|c|c|  }
		\multicolumn{4}{c}{} \\ 
		$\varepsilon_s$ &$\varepsilon_\infty$& $\tau_e$& $\sigma_e$& $\mu_{s}$& $\mu_\infty$& $\tau_m$& Type \\ 
		\hline
		15	& 5.5&0.02 ns& 0.1 S/m& 4& 1.5& 0.05ns& B1  \\
		
	\end{tabular}
	
	\begin{tabular}{ c|c|c|c|c|c|c|c|c|c|  }
		\multicolumn{4}{c}{} \\ 
		$\varepsilon_s$ &$\varepsilon_\infty$& $\delta_e$& $f_{0,e}$& $\sigma_e$&$\mu_{s}$& $\mu_\infty$& $\delta_m$&$f_{0,m}$& Type \\ 
		\hline
		2	& 1.5&9.5GHz& 10.5GHz&0.4 S/m& 3& 2.5& 10.5GHz& 9GHz& B2 \\
	\end{tabular}
\end{table}

\section{Possible Realization Schemes and Uncertainty Analysis}
In this section, the feasibility of the proposed time-varying characterization method is investigated via presenting two different electrical and mechanical tuning mechanisms. The whole setup is composed of four main parts: 1) waveguide transitions, measurement cell, adapters, and the other accessories, 2) external time-varying actuator along with peripheral programming circuits, 3) signal generator, spectrum analyzer, and auxiliary devices such as directional coupler, and 4) data acquisition and post-processor systems. As can be seen from \textcolor{blue}{Fig. 7}, the MUT is embedded inside the X-band measurement cell (a non-reflecting waveguide section). The waveguide transition with length $>$ 70 mm is in charge of filtering out the higher-order modes, if any,  between the measurement cell and adapter. A standard directional coupler with known coupling ($C$), isolation ($I$), and directivity ($D$) coefficients is employed to transmit the incident signals from the signal generator to the coaxial-to-waveguide adapter and also, a certain fraction of the reflected harmonic components to the spectrum analyzer. Signal generator, spectrum analyzer, and  coaxial-to-waveguide adapter are connected to ports 3, 1, and 2, respectively, while port 4 is terminated by a match load. The effect of the microwave coupler can be simply diminished by using a simple amplitude calibration. The spectrum analyzer reads the power of the emerging harmonics, ${P}_{\text{reflected,m}}$, and sends them to the post-processing system for extraction of $\varepsilon_r\left(\omega\right)$=$\varepsilon_r^\prime\left(\omega\right)-j\varepsilon_r^{\prime\prime}\left(\omega\right)$ and $\mu_r\left(\omega\right)$=$\mu_r^\prime\left(\omega\right)-j\mu_r^{\prime\prime}\left(\omega\right)$. The amplitude of each harmonic can be calculated by   
\begin{align}
& \left| {{a}_{m}} \right|=\sqrt{\frac{{{P}_{\text{reflected,m}}}}{{{P}_{\text{reference}}}}}
\end{align}
provided that high isolation (I$>$40 dB) between the signal generator and spectrum analyzer is supplied. ${P}_{\text{reference}}$ is the power read by the spectrum analyzer when no MUT is placed within the short-circuited measurement cell.
 
The temporal modulation can be realized with either electrical or mechanical mechanism. Under mechanical reconfiguration, a movable short-circuit plate terminates the measurement cell. The accurate displacement of the metal plate can be guaranteed by a stepper motor, a rotatable cylinder, and a non-rotatable square-section shaft. The working principle for precise control over the position of the short-circuit terminations is as follows. The stepper motor is driven by a power driver that is connected to an external micro control unit and rotates the rotatable cylinder. The rotatable shaft is actuated by clockwise/counter-clockwise rotation of the rotatable cylinder and pushes/pulls the non-rotatable square-section shaft together with the metal plate on the top through a soft joining mechanical structure. The switching program is already saved into the memory of micro control unit. Given the maximum available speed of the linear actuators up to 15000 mm/s, the frequency gaps between the neighboring harmonics can be chosen up to 100 Hz. Despite being very small in comparison to the central frequency, the frequency intervals are still high enough to be discriminated by the spectrum analyzers. 

According to \textcolor{blue}{Eq. (6)}, the short-circuited air-filled waveguide section shown in \textcolor{blue}{Fig. 1} can be circuitally modeled as a purely imaginary impedance or a termination load with $\Gamma_L$=exp$(j\theta)$. As an alternative route to make a change in the physical location of the short-circuit termination, the metal plate movement can be equivalently carried out by a real-time electrical control over the reflection phase $\theta$(t)=2Im$\left\{ {{\gamma }_{z0}}L\left( t \right) \right\}$. Owing to the recent developments of microwave metasurfaces and reflectarrays with programmable reflective meta-atoms spanning the entire 0$-$2$\pi$ phase range \cite{Iqbal, Yang,Trampler,Costanzo}, the dual- or triple-state time modulation of reflection phase is quite feasible and lies within the realm of current fabrication technologies. In this case, the short-circuit termination is replaced with a group of sub-wavelength phase-agile meta-atoms. The varactor diodes with controllable DC biasing voltages are embedded into the meta-atoms, providing the frequency modulation up to several MHz \cite{He}. The biasing voltages can be produced by an external micro control unit in which the reflection phase sequences are already stored. Although the electrical switch gives a faster modulation which, in turn, leads to a better frequency discrimination in the spectrum analyzer, the slight losses associated with the meta-atoms may cause small tolerances in the reconstructed $\varepsilon_r\left(\omega\right)$=$\varepsilon_r^\prime\left(\omega\right)-j\varepsilon_r^{\prime\prime}\left(\omega\right)$ and $\mu_r\left(\omega\right)$=$\mu_r^\prime\left(\omega\right)-j\mu_r^{\prime\prime}\left(\omega\right)$.

In the end, we want to analyze the sensitivity of the proposed algorithm with respect to any uncertainty that may occur in our measurements. It is to be underlined that there are many mechanisms contributing to the error budget in reconstruction of the permittivity and permeability parameters. 
Since there are already many works related to the usual uncertainty terms (arising from higher-order modes, lateral fitting, sample thickness, and so on) \cite{Hasar2009,Hasar20092,Hasar2015,Jarvis}, we focus on the uncertainty in measuring the harmonic intensities (originated from the unwanted loss of axillary meta-atoms, the ambient noise, and so on) and setting the position of short-circuit termination (or equivalently, the reflection phase of the auxiliary meta-atoms). The former case is studied by corrupting the synthetic data of the harmonic intensities with Additive White Gaussian Noise (AWGN) with a mean value equal to zero and a specific signal-to-noise ratio (SNR). Without loss of generality, the sensitivity analysis is accomplished for a test specimen with $\varepsilon_r$=4.5, $\mu_r$=2.5, $\tan \delta_e$=0.05, and thickness $h_1$=3 mm. The reconstruction algorithm is applied to the noisy data and the corresponding results are displayed in \textcolor{blue}{Fig. 8a} for different SNR values. We proceed with the effect of uncertainty in the position of short-circuit termination. Two different cases are considered here in one of which $L_1$, $L_2$, and $L_3$ are slightly offset from their actual values, while in the other one, only $L_1$ has experienced a small misalignment. For different values of shift percentage (${\Delta L}/{L}\;$), the retrieved parameters of the MUT are reported in \textcolor{blue}{Fig. 8b}. With an overall view on the results, it may be concluded that for both cases, the amount of discrepancies between the actual and extracted data tends to increase as the uncertainties get worse. However, for the position tolerances up to 5\% and SNR values exceeding 14 dB, respectively, the errors in $\varepsilon_r\left(\omega\right)$ and $\mu_r\left(\omega\right)$ determination are not much significant ($<$5\%).

\begin{figure}[t]
	\includegraphics[scale=0.3]{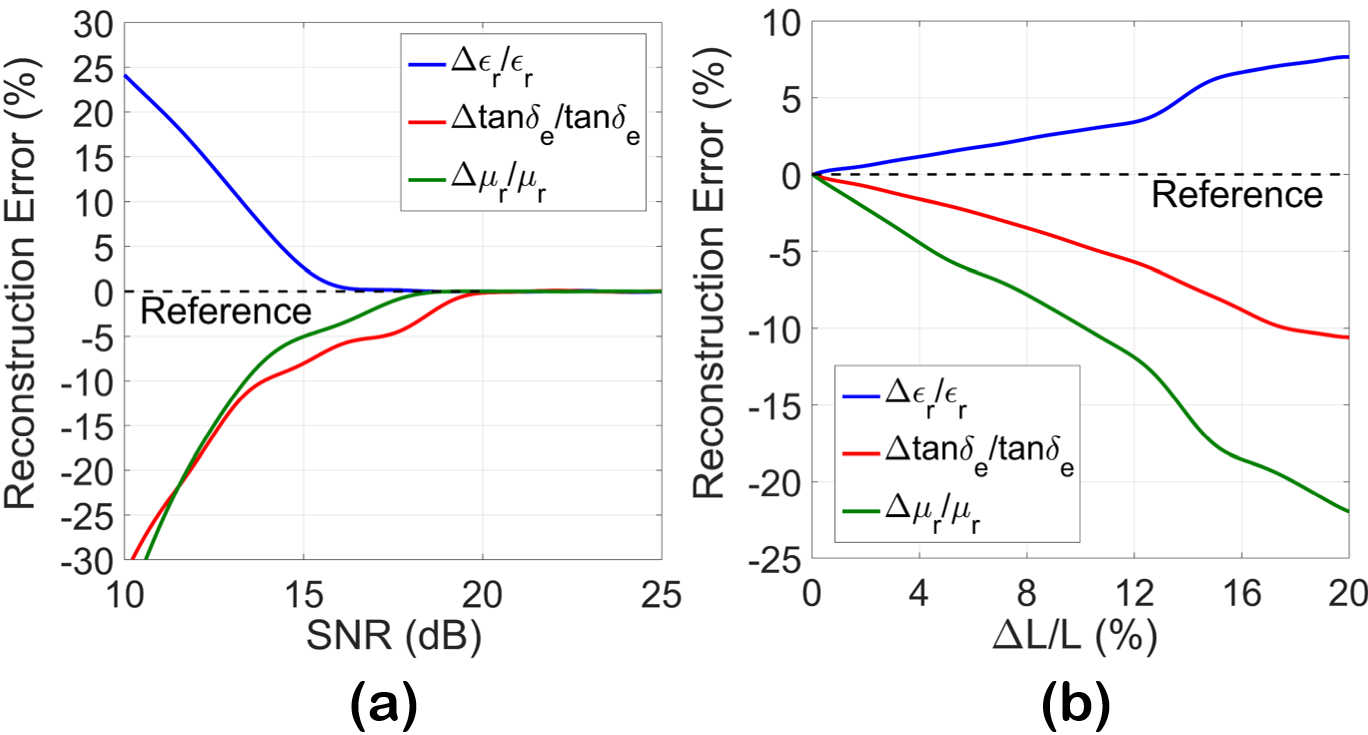}
	\caption{\label{fig:epsart} {Reconstruction error in dielectric properties of a test specimen with $\varepsilon_r$=4.5, $\mu_r$=2.5, $\tan \delta_e$=0.05, and thickness $h_1$=3 mm \& $h_2$=1 mm due to (a) position misalignment and (b) noise.} }
\end{figure}

\section{conclusion}
In summary, we proposed an analytical strategy based on temporal-modulation of short-circuited waveguides for unique extraction of the electromagnetic parameters of MUT with amplitude-only reflection measurements. We demonstrated that when the position of short-circuit termination is switched cyclically in a predefined way, a set of harmonic components emerge whose intensities are a nonlinear function of the reflection phases. Several examples were illustrated for which the true and reconstructed electromagnetic parameters of the lossy dispersive specimens are in an excellent agreement without any priori knowledge of MUT parameters. The feasibility of the proposed approach was comprehensively discussed through two different realization schemes and the uncertainty analysis. We note that even though our demonstrations are presented for X-band waveguide setup, the proposed material characterization method can also be applied for sheet resistance extraction \cite{Feng} and metasurface susceptibility derivation \cite{momeni}, complex (anisotropic, inhomogeneous, and chiral) materials \cite{Kiani,bahar,Zarifi}, and extended to free-space configurations and even, acoustic platforms. The proposed concept sets a new stage for $\varepsilon_r$ and $\mu_r$ determination in programmable time-modulated waveguides and may find the other potential applications such as efficient harmonic control for guided waves. \\ \\

\section{APPENDIX}
The harmonic components $a_m$ as the Fourier series coefficients of $S_{11}$(t) are derived in detail as follows: 
\begin{align}
\nonumber{{a}_{m}}=\frac{1}{{{T}_{m}}}\int\limits_{{{T}_{m}}}{{{S}_{11}}\left( t \right)\exp \left( -j2\pi m{{f}_{m}}t \right)dt} \\ \nonumber=\frac{1}{{{T}_{m}}}\sum\limits_{n=1}^{q}{{{S}_{11,n}}\int\limits_{{{T}_{m}}}{{{\Pi }_{n}}\left( t \right)\exp \left( -j2\pi m{{f}_{m}}t \right)dt}} \\ 
\nonumber =\frac{1}{{{T}_{m}}}\sum\limits_{n=1}^{q}{{{S}_{11,n}}\int\limits_{\left( n-1 \right){{T}_{m}}/q\;}^{n{{T}_{m}}/q\;}{\exp \left( -j2\pi m{{f}_{m}}t \right)dt}}\\
\nonumber=\sum\limits_{n=1}^{q}{\frac{{{S}_{11,n}}}{j2m\pi }}\left[ \exp \left( -j2\pi m\frac{n-1}{q} \right)-\exp \left( -j2\pi m\frac{n}{q} \right) \right]\\
\nonumber=\frac{1}{m\pi }\sum\limits_{n=1}^{q}{{{S}_{11,n}}}\sin \left( \frac{m\pi }{q} \right)\exp \left[ -j\frac{m\pi \left( 2n-1 \right)}{q} \right] \tag{A1}
\end{align}

In addition, the detail expressions for the theoretical values of $r$ and $\Delta \phi$ are 
\begin{widetext}
	\begin{align}
  {{r}^{\text{TH}}_{n}}=\left| \frac{Z_{0}^{TE}\tanh \left( {{\gamma }_{z0}}{{L}_{n}} \right)+Z_{d}^{TE}\tanh \left( {{\gamma }_{zd}}h \right)-{{\left( Z_{0}^{TE} \right)}^{2}}\tanh \left( {{\gamma }_{z0}}{{L}_{n}} \right)\tanh \left( {{\gamma }_{zd}}h \right)-Z_{0}^{TE}Z_{d}^{TE}}{Z_{0}^{TE}\tanh \left( {{\gamma }_{z0}}{{L}_{n}} \right)+Z_{d}^{TE}\tanh \left( {{\gamma }_{zd}}h \right)+{{\left( Z_{0}^{TE} \right)}^{2}}\tanh \left( {{\gamma }_{z0}}{{L}_{n}} \right)\tanh \left( {{\gamma }_{zd}}h \right)+Z_{0}^{TE}Z_{d}^{TE}} \right| \tag{A2} \\
  \Delta {{\phi }^{\text{TH}}_{ij}}=\measuredangle \frac{Z_{0}^{TE}\tanh \left( {{\gamma }_{z0}}{{L}_{i}} \right)+Z_{d}^{TE}\tanh \left( {{\gamma }_{zd}}h \right)-{{\left( Z_{0}^{TE} \right)}^{2}}\tanh \left( {{\gamma }_{z0}}{{L}_{i}} \right)\tanh \left( {{\gamma }_{zd}}h \right)-Z_{0}^{TE}Z_{d}^{TE}}{Z_{0}^{TE}\tanh \left( {{\gamma }_{z0}}{{L}_{j}} \right)+Z_{d}^{TE}\tanh \left( {{\gamma }_{zd}}h \right)-{{\left( Z_{0}^{TE} \right)}^{2}}\tanh \left( {{\gamma }_{z0}}{{L}_{j}} \right)\tanh \left( {{\gamma }_{zd}}h \right)-Z_{0}^{TE}Z_{d}^{TE}} \times \tag{A3} \\
  \nonumber \frac{Z_{0}^{TE}\tanh \left( {{\gamma }_{z0}}{{L}_{j}} \right)+Z_{d}^{TE}\tanh \left( {{\gamma }_{zd}}h \right)+{{\left( Z_{0}^{TE} \right)}^{2}}\tanh \left( {{\gamma }_{z0}}{{L}_{j}} \right)\tanh \left( {{\gamma }_{zd}}h \right)+Z_{0}^{TE}Z_{d}^{TE}}{Z_{0}^{TE}\tanh \left( {{\gamma }_{z0}}{{L}_{i}} \right)+Z_{d}^{TE}\tanh \left( {{\gamma }_{zd}}h \right)+{{\left( Z_{0}^{TE} \right)}^{2}}\tanh \left( {{\gamma }_{z0}}{{L}_{i}} \right)\tanh \left( {{\gamma }_{zd}}h \right)+Z_{0}^{TE}Z_{d}^{TE}}
	\end{align}
\end{widetext}

%\nocite{*}

\end{document}